\documentclass[letterpaper]{article} % DO NOT CHANGE THIS
\usepackage{aaai2026} % DO NOT CHANGE THIS
\usepackage{times}  % DO NOT CHANGE THIS
\usepackage{helvet}  % DO NOT CHANGE THIS
\usepackage{courier}  % DO NOT CHANGE THIS
\usepackage[hyphens]{url}  % DO NOT CHANGE THIS
\usepackage{graphicx} % DO NOT CHANGE THIS
\usepackage{natbib}  % DO NOT CHANGE THIS AND DO NOT ADD ANY OPTIONS TO IT
\usepackage{caption} % DO NOT CHANGE THIS AND DO NOT ADD ANY OPTIONS TO IT
\usepackage{algorithm}
\usepackage{algorithmic}

\usepackage{newfloat}
\usepackage{listings}
\DeclareCaptionStyle{ruled}{labelfont=normalfont,labelsep=colon,strut=off} % DO NOT CHANGE THIS
\floatstyle{ruled}
\newfloat{listing}{tb}{lst}{}
\floatname{listing}{Listing}
\newcommand{\sys}{{LOOM}{}}

\title{LOOM: Personalized Learning Informed by Daily LLM Conversations Toward Long-Term Mastery via a Dynamic Learner Memory Graph}

\author{
    Justin Cui, Kevin Pu, Tovi Grossman
}

\affiliations{
    University of Toronto, Toronto, Ontario, Canada\\
    justin.cui@mail.utoronto.ca, jpu@dgp.toronto.edu,
    tovi@dgp.toronto.edu
}

\begin{document}

\maketitle

\begin{abstract}
    % AI technologies have influenced how learning systems are built and how users interact with them.
    % Many existing tools focus on using AI to personalize a pre-defined set of learning material or adapt to user learning progress. 
    % However, many systems still face lack of adoption due to rigid learning plan and a lack of understanding to the user's learning needs and capacity to learn in their daily lives.
    % From a formative study, we identified three main learning needs and built \sys{}, a personalized micro-learning application that observes the user's learning needs in daily search queries and proactively generates micro-learning lessons that adapts to the user's capacity of that day.
    % \sys{} also uses a knowledge graph to systematically build up the user's understanding around the topic over time.
    % Through a user study with X participants, we found that \sys{} more closely aligned with users' learning needs, was more engaging, and resulted in similar learning outcomes compared to a baseline system.
    % We discuss how our pipeline can adapt to different learning goals and domains, and propose design implications to leverage user context to provide a truly personalized learning experience.
    Foundation models are increasingly used to personalize learning, yet many systems still assume fixed curricula or coarse progress signals, limiting alignment with learners’ day-to-day needs. At the other extreme, lightweight incidental systems offer flexible, in-the-moment content but rarely guide learners toward mastery. Prior work privileges either continuity (maintaining a plan across sessions) or initiative (reacting to the moment), not both, leaving learners to navigate the trade-off between recency and trajectory—immediate relevance versus cumulative, goal-aligned progress. 
    We present LOOM, an agentic pipeline that infers evolving learner needs from recent LLM conversations and a dynamic learner memory graph, then assembles coherent learning materials personalized to the learner’s current needs, priorities, and understanding. These materials link adjacent concepts and surface gaps as tightly scoped modules that cumulatively advance broader goals, providing guidance and sustained progress while remaining responsive to new interests. 
    We describe LOOM’s end-to-end architecture and working prototype, including conversation summarization, topic planning, course generation, and graph-based progress tracking. In a formative study with ten participants, users reported that LOOM’s generated lessons felt relevant to their recent activities and helped them recognize knowledge gaps, though they also highlighted needs for greater consistency and control. We conclude with design implications for more robust, mixed-initiative learning pipelines that integrate structured learner modeling with everyday LLM interactions.
\end{abstract}


\section{Introduction}

General-purpose chat assistants like ChatGPT are good at answering explicit questions, but they are not designed to proactively guide learning tailored to each user's specific needs \cite{proactive2023liao, wang2025enhancinguserorientedproactivityopendomain}. While general-purpose chat assistants typically accumulate rich context from frequent user interactions, they rarely leverage this proactively to structure personalized learning or help learners recognize their own “unknown unknowns”—knowledge gaps they sense but cannot explicitly name. This leaves learners bearing the burden of identifying next steps and integrating isolated insights into a coherent, sustained learning trajectory. The result is user-initiated assistance that feels useful in the moment yet fragmented and disconnected over time.

LLM-based learning tools already deliver on the \textit{pedagogy} pillar—they explain, quiz, and give feedback \cite{chattutor2024chen, ye2025positionllmsgoodtutors}, but in practice they tend to emphasize either \textbf{continuity} or \textbf{initiative/flexibility}, rarely both together. Systems that prioritize continuity maintain a plan or learner model across sessions, yet usually assume declared goals and update mainly inside the tutoring context \cite{chattutor2024chen, li2025tutorllmcustomizinglearningrecommendations}; they seldom take the initiative to read a learner’s ongoing activity (e.g., everyday queries/chats) and propose adaptive, goal-aligned next steps when objectives are hazy under “unknown unknowns.” Conversely, tools that emphasize initiative and flexibility are responsive to the moment and good at helping learners start \cite{aiget2025cai, Arakawa2022VocabEncounterNV, micromandarin2011edge}, but they rarely carry forward those micro-steps into a durable trajectory that tracks mastery over days or weeks. This fragmentation places the cognitive burden on learners, forcing them to manually integrate immediate interactions into a meaningful, long-term trajectory. This underscores the need for a unified approach that seamlessly merges continuity, initiative, and pedagogy.

Prior LLM-based learning tools often fall into three threads: (1) \textbf{chat-based and target-focused tutors} \cite{chattutor2024chen, li2025tutorllmcustomizinglearningrecommendations, Park_2024} that often track learning paths along a set trajectory over time, yet don’t watch users’ everyday activity; they adapt in relation to the pre-defined curriculum, not to evolving information needs or daily time/attention; (2) \textbf{incidental learning systems} \cite{aiget2025cai,Arakawa2022VocabEncounterNV,micromandarin2011edge, waitchatter2015cai} that surface in-situ content but do not scaffold these experiences toward durable, long-term understanding; and (3) \textbf{memory-augmented assistants} \cite{gum2025shaikh, baek2024knowledgeaugmentedlargelanguagemodels, Zulfikar_2024, omniquery2024li} that keep rich personal context yet are not pedagogical as they don’t track mastery or assemble lessons into a coherent progression.

To bridge these gaps, we present \textbf{LOOM}\footnote{We provide an anonymous GitHub repo with the source code and a video demo here: https://anonymous.4open.science/r/LoomDemo}, a personalized learning system that observes everyday LLM conversations, infers evolving needs, and assembles coherent learning modules scoped to each learner’s current priorities and understanding. At its core is a dynamic learner memory graph that tracks mastery, links adjacent concepts, and continuously updates based on new interactions. LOOM unifies \textit{continuity} and \textit{initiative} in a single agentic pipeline, maintaining long-term progression while remaining responsive to spontaneous learning opportunities.

% Our research questions are: \textit{\textbf{RQ1}}: Can personalized learning driven by everyday LLM conversations better align with learners’ needs and boost engagement compared to a reactive chat baseline? \textit{\textbf{RQ2}}: Can \sys{} achieve comparable learning outcomes while remaining responsive to evolving goals and interests? \textit{\textbf{RQ3}}: How does a dynamic learner memory graph influence learners’ sense of coherence and progress over time?

We conducted a formative study with ten participants who used LOOM over two days alongside their normal LLM-based activities. Participants explored the generated mini-courses and reflected on their perceived relevance and coherence through a post-study survey. Results show that learners found the generated lessons closely aligned with their recent conversations and helpful for recognizing knowledge gaps, while also pointing to needs for stronger consistency and greater control over pacing and content.

We summarize our contributions as follows:
\begin{enumerate}
    \item \textbf{System design and prototype.} We introduce \sys{}, a personalized learning system underpinned by a dynamic learner memory graph, and describe its end-to-end pipeline for inferring goals, planning adaptive lessons, and tracking mastery through everyday LLM interactions.
    \item \textbf{Formative user evaluation.} We report early findings from a ten-participant study highlighting \sys{}’s perceived relevance, coherence with prior chats, and its ability to surface “unknown unknowns,” along with areas for improvement in content depth and pacing.
    \item \textbf{Design implications and future directions.} We discuss directions for enhancing the robustness of the agentic pipeline and enabling mixed-initiative learner interaction through editable, visual learner graphs that balance system initiative with learner agency.
\end{enumerate}

% --- LOOM system diagram: full-width (two-column) ---
\begin{figure*}[!t]      % stronger placement hint; still a float and can only go top-of-page
  \centering
  \includegraphics[width=\textwidth]{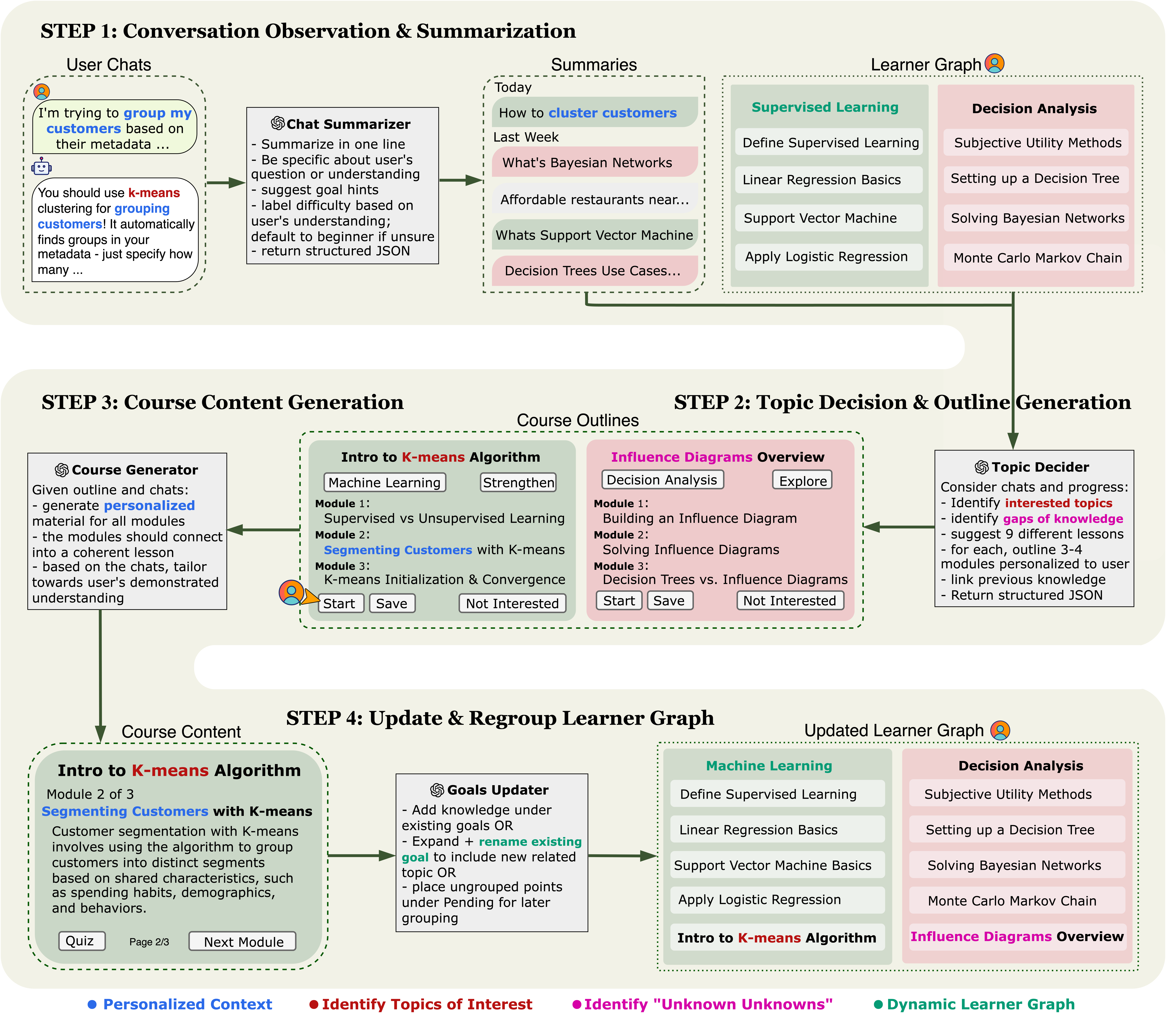}
  \caption{The \sys{} pipeline unifies initiative and continuity for personalized learning: 
  \textbf{Step 1}: \textbf{Conversation Observation \& Summarization} surfaces recurring themes and gaps in user's chats with an LLM assistant; 
  A \textit{learner graph} organizes mentioned concepts and demonstrated prior knowledge and learning progress; 
  \textbf{Step 2}: \textbf{Topic Decision \& Outline Generation} identifies learning topic from user inquiries and potential adjacent concept to explore, displayed as course outline proposed to the user;
  \textbf{Step 3}: \textbf{Course Content Generation} serves adaptive modules linked to adjacent concepts and current goals; 
  \textbf{Step 4}: \textbf{Updated \& Regroup Learner Graph} records user engagement and learning outcomes to update mastery progress and guide next steps and reinforcement over time.}
  \label{fig:loom-system}
\end{figure*}

\section{Related Work}

\subsection{General-Purpose Chat Assistants}
Large-scale conversational agents such as ChatGPT and its successors have been widely adopted as on-demand learning aids, given their broad knowledge and fluent dialogue capabilities \cite{Shen2023ChatGPTAO, xu2024largelanguagemodelseducation}. These assistants can answer questions across domains and have been embedded in educational platforms. For example, Khan Academy’s \textit{Khanmigo} (built on GPT-4) acts as a cross-subject tutor engaging students in problem solving and role-playing scenarios \cite{khanmigo}. Duolingo’s GPT-4 integration enables open-ended conversational practice and real-time explanations beyond scripted lessons \cite{duolingo}.

While these systems demonstrate the pedagogical potential of foundation models, they remain largely \textit{reactive}—responding to user prompts without maintaining a persistent model of the learner’s goals or progress. They rarely intervene proactively or scaffold cumulative understanding, motivating specialized approaches that can align with a learner’s evolving context.

\subsection{Structured, Target-Focused Learning}
A complementary body of work builds structured tutoring systems around LLMs. These systems assume explicit learning goals (e.g., “learn Python,” “prepare for exam X”) and personalize instruction, practice, and assessment toward them.

ChatTutor chains multiple LLM roles to handle instruction, reflection on performance, and adaptive quizzes and course adjustments \cite{chattutor2024chen}. TutorLLM combines learner modeling with retrieval-augmented generation and knowledge tracing to recommend next steps based on predicted mastery \cite{li2025tutorllmcustomizinglearningrecommendations}. Conversation-based tutors that explicitly model student knowledge similarly tailor questions and feedback across a session \cite{Park_2024}.

These systems demonstrate strong \textit{goal-aligned continuity}: they maintain learner models, track progress, and adapt within well-defined sessions. Yet they typically rely on declared goals and assume dedicated study time, making them less responsive to the fluid, opportunistic nature of everyday learning. They seldom detect new interests or “unknown unknowns” emerging from spontaneous activity.

\subsection{Incidental Learning Systems}
Incidental learning systems embed micro-lessons into daily routines, transforming idle moments into learning opportunities. Early work like WaitChatter introduced second-language vocabulary practice into instant-messaging pauses \cite{waitchatter2015cai}, while MicroMandarin surfaced location-relevant phrases during real-world interactions \cite{micromandarin2011edge}. More recent systems such as VocabEncounter automatically generate contextual examples for newly learned words and resurface them across device use \cite{Arakawa2022VocabEncounterNV}.

Recent embodied extensions, such as AiGet on smart glasses, observe users’ surroundings to infer interesting objects or concepts and proactively deliver short contextual explanations \cite{aiget2025cai}. These systems exemplify \textit{initiative and timeliness}: they notice teachable moments and provide just-in-time content. However, they rarely maintain long-term learner models or connect individual encounters into a coherent progression, leaving mastery untracked over time.

\subsection{Memory-Augmented Personal Assistants}
Another relevant direction explores assistants that maintain persistent user context across interactions. Rather than treating each chat independently, these systems build long-term representations of the user’s activities and preferences. For instance, Shaikh et al.\ learn a General User Model (GUM) from everyday computer use to anticipate needs and act proactively \cite{gum2025shaikh}. Other approaches augment LLMs with persistent memory stores, enabling retrieval of prior conversations and personalization based on episodic recall \cite{baek2024knowledgeaugmentedlargelanguagemodels, Zulfikar_2024}. Work on proactive conversational agents likewise argues that such memory is foundational for assistants that suggest next steps rather than waiting for explicit queries \cite{proactive2023liao}.

These systems achieve continuity of personal context—they remember who the user is and what they have done—but they are not explicitly pedagogical. They optimize for productivity or information access, not for tracking mastery or constructing cumulative learning trajectories \cite{gum2025shaikh, baek2024knowledgeaugmentedlargelanguagemodels, Zulfikar_2024}.

\section{System Implementation}

\subsection{Overview}
We introduce LOOM, a system that aims to resolve the tension between continuity and initiative through a four-stage pipeline that transforms everyday conversations with LLM-based tools like ChatGPT into personalized learning trajectories (Fig.\ref{fig:loom-system}). Unlike traditional intelligent tutoring systems that require explicit educational intent upfront, or incidental learning tools limited to transient content, LOOM proactively identifies latent learning needs directly from learners' everyday interactions. By continuously monitoring the user’s LLM conversations, LOOM infers learner needs from the questions they ask, topics they explore, and misconceptions they reveal. At the pipeline's core is a dynamic learner memory graph—a structured representation of the learner’s evolving knowledge, tracking learning goals and knowledge connections over time.
The pipeline consists of four interconnected stages: (1) conversation observation and summarization, extracting concise learning signals from everyday chats; (2) topic decision and course outline proposal, identifying timely learning opportunities aligned with recent interests and existing knowledge; (3) course content generation, producing personalized content tailored to the learner's prior interactions; and (4) progress tracking and memory graph updates, refining the learner model over time. Each stage is implemented as a lightweight LLM-based agent with defined inputs and outputs, making the pipeline modular and extensible for future rule-based or retrieval-augmented improvements.

\subsection{Conversation Observation and Summarization}
To address the limitation that structured learning systems rarely proactively monitor learners' ongoing activities, LOOM continuously observes learner conversations with an LLM assistant. The assistant utilizes a similar interface with popular chat-based AI tools like ChatGPT and Gemini, where users can enter queries and keep track of history in a left panel. The observation layer distills conversational exchanges into concise learning signals, capturing what the user was trying to learn without requiring explicit educational intent.
For each conversation, the system invokes a summarizer agent to generate a single, specific, learner-centric statement capturing exactly what the user discussed or asked. 
For example, a user's query to ask for help in grouping customer metadata might lead to a conversation with LLM discussing grouping using K-means clustering, which is summarized as “How to cluster customers” (Fig.\ref{fig:loom-system}, Step 1).
Additionally, each summary tags the conversation with a thematic umbrella (e.g., “Supervised Learning”) to guide future educational topic suggestions, along with an estimated difficulty level for the user (beginner, intermediate, or advanced) based on the user's demonstrated understanding in the way they discussed the topic with the LLM.
To ensure relevance, the system maintains an activity-based filter: conversations are marked as “active" if referenced recently (e.g., within 10 days). This temporal filtering ensures course recommendations reflect current priorities, addressing critiques of static curricula that fail to adapt to evolving interests.

\subsection{Topic Decision and Course Outline Proposal}
The topic decision stage integrates recent conversation summaries with the learner memory graph to identify coherent, timely learning opportunities grounded in both recent activity and accumulated progress. Specifically, the system utilizes a topic decider agent to select themes frequently appearing across recent interactions or topics closely related to the learner's existing knowledge, ensuring that suggested courses are either clearly tied to previously discussed topics (strengthen mode) or introduce relevant adjacent concepts (explore mode). 
For example, \sys{} identifies the outline for K-means algorithm based on previously discussed topics, and proactively introduces a course outline for influence diagram since it is an adjacent concept to the user's learner memory graph on decision analysis, surfacing a potential “unknown unknown" based on existing knowledge (Fig.\ref{fig:loom-system}, Step 2).
The selection process prioritizes both immediate relevance and long-term guidance, ensuring suggested courses align with the learner’s evolving interests and broader mastery goals.
For each identified learning topic, the system proposes concise mini-course outlines consisting of 3–4 modules, each clearly linked to recent interactions through specific source chats. Proposals include a succinct goal label to help learners quickly scan suggestions, a concise course title, and a few learner-facing questions the course will address. Each module is structured with explicit time estimates, facilitating manageable, targeted learning sessions. This mixed-signal approach allows LOOM to surface both reinforcement topics that consolidate prior knowledge and adjacent “unknown-unknowns” that expand the learner’s conceptual graph. This design also ensures that each course feels both immediately relevant and personally tailored to the learner’s recent activity and goals, fostering a coherent and engaging learning experience.

\subsection{Dynamic Learner Memory Graph Construction}
The learner memory graph maintains continuity across learning sessions by representing the learner’s knowledge landscape hierarchically into clearly labeled goal umbrellas and their associated courses.
Goals are broad thematic categories (e.g., “Machine Learning" or “Decision Analysis") dynamically adjusted as learners complete courses. Goals primarily track completed courses as concrete evidence of mastery, enabling dynamic regrouping of new material under evolving thematic umbrellas.
Courses within the graph explicitly connect to these high-level goals. Each course is associated with a goal and maintains concise module-level tracking to clearly represent learner progress. This structured representation distinguishes LOOM from general memory-augmented assistants by explicitly supporting coherent, pedagogically meaningful learning trajectories. In our prototype, we use a simple two-level graph that links goals to their courses and tracks per‑module progress on each course, keeping the representation lightweight and easy to extend. This representation keeps the model interpretable for learners and provides a clear structure for visualizing progress within the interface.

\subsection{Full Course Generation}
Once a learner selects a proposed outline, LOOM runs a content generation agent that converts relevant recent conversations into a personalized, coherent mini-course. The system retrieves the flagged source conversations, extracts actual chat excerpts that capture the learner’s phrasing and intent, and pairs them with the selected outline to guide generation. An LLM then produces content for the planned 3–4 modules tailored to the learner’s demonstrated level and language: each module contains concise lesson text and a small multiple-choice quiz to diagnose understanding and surface remaining gaps (Fig.\ref{fig:loom-system}, Step 3). 
The generation agent enforces lightweight pedagogical constraints, sequencing from core ideas to applications and synthesis to maintain coherence across modules.
Grounding modules in the learner’s own dialogue keeps each mini‑course immediately personalized while successive courses accumulate toward broader, goal‑aligned mastery.

\subsection{Progress Tracking and Learner Graph Updates}
To maintain a precise learner model, LOOM continually updates the memory graph based on actual learner engagement.
The system records module-level course completions explicitly when learners actively finish lessons and quizzes. Learners can also self-report mastery through the outline’s “Questions you will answer” interface by marking modules as already known or irrelevant, so the graph more accurately reflects true understanding.
A crucial part of our pipeline is the dynamic regrouping of completed courses into broader goal umbrellas in the learner graph. After each completion, a regrouping agent proposes structured updates: adding courses to existing goals, renaming goal umbrellas when broader themes emerge, or creating a new goal only when multiple related courses form a coherent cluster.
For example, after engaging with course modules on K-means and influence diagrams, the learner graph re-frames the first goal from “Supervised Learning" to “Machine Learning" to account for the added learning on the K-means algorithm. Meanwhile, the influence diagrams course fits well under decision analysis and is updated as well (Fig.\ref{fig:loom-system}, Step 4).
Each regrouping action is concise, clear, and directly tied to recent learner activity, ensuring the learner memory graph remains pedagogically meaningful and coherent as learner expertise evolves.
These updates immediately adjust user goals to reflect updated interests and progress, and steer the next set of suggestions to match current priorities.
By continuously refining this structured representation based strictly on concrete learner completions, LOOM maintains a robust and adaptive mastery model, effectively balancing immediate responsiveness with long-term guidance to structured, cumulative knowledge growth.

\section{Evaluation}

\subsection{Method}
\textbf{Study design:} We ran a formative study with ten participants to gather early signals about LOOM’s usefulness, coherence, and capability to surface gaps from everyday chats. Participants consented to the study and were asked to use LOOM over two days and to carry out their normal conversational workflows: specifically, each participant was instructed to create at least 10--15 new chats that reflected real questions they would ask in work, study, or other scenarios in their daily workflow. After producing these chats, participants engaged with suggested mini-courses (outlines, lessons, quizzes). The system automatically observed these sessions and generated personalized lesson proposals via its four-agent pipeline. After the study, participants completed a survey with Likert-scale questions and open-text prompts. All participant data are de-identified and encrypted before storage and analysis. 

\textbf{Participants:} Ten participants completed the study (convenience sample drawn from colleagues and professional contacts), denoted as P1-P10. All participants finished the quantitative survey; nine provided free-text qualitative feedback.

\textbf{Measures:} We collected a 10-item Likert questionnaire (1 = Strongly Disagree, 7 = Strongly Agree) covering usefulness, repetition, consolidation, novelty, gap discovery, length/depth, coherence, trust, motivation, and likelihood of reuse. Two open-ended prompts asked (1) what participants liked most about LOOM and (2) what they would change to improve the suggested topics and learning materials.

\subsection{Results}

\subsubsection{Quantitative summary:} 
\begin{figure*}[t]
\centering
\includegraphics[width=\textwidth]{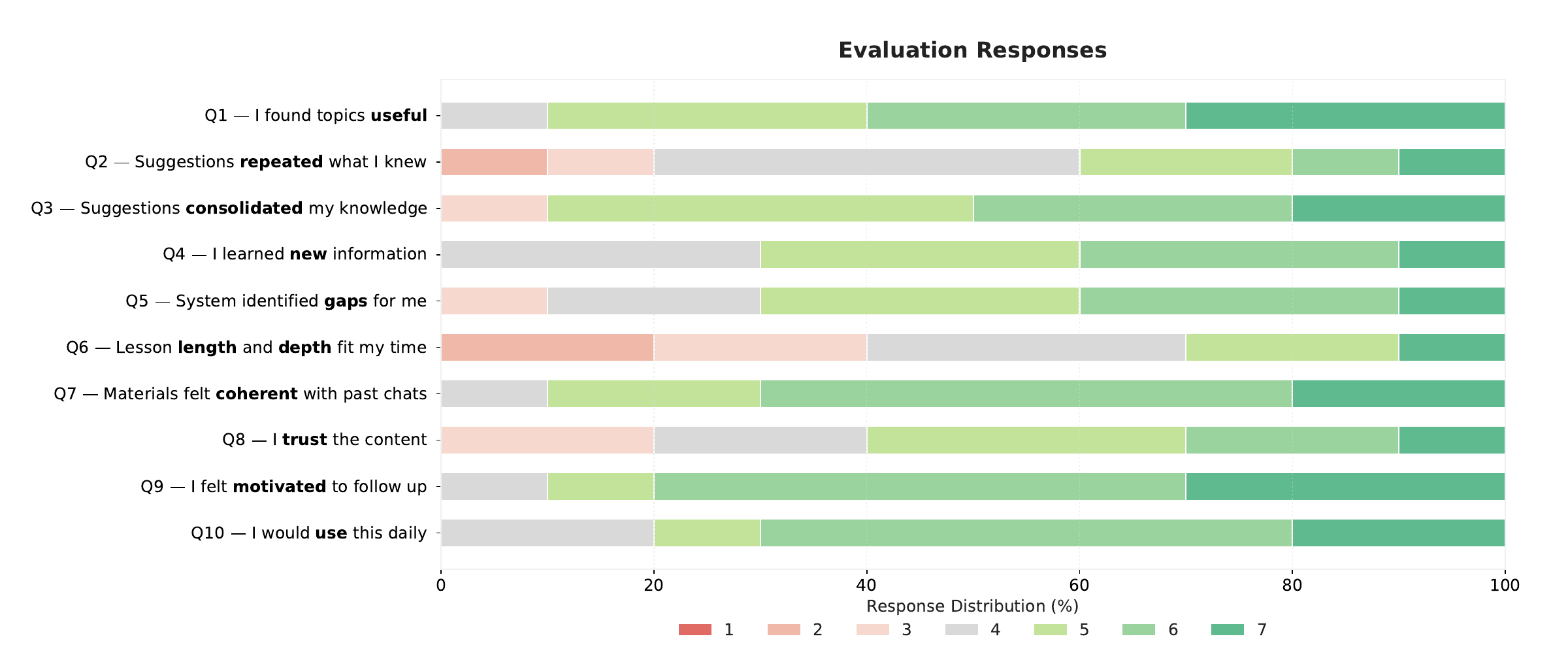}
\caption{Evaluation responses: stacked distributions of participants' Likert responses (colors indicate response bands from Strongly Disagree to Strongly Agree). Q2 is reverse-coded, so higher scores indicate fewer repetitions (a more positive outcome).}
\label{fig:eval_responses}
\end{figure*}

Figure~\ref{fig:eval_responses} shows the full response distributions for each survey item (Q1–Q10). Two clear patterns are visible. First, responses skew positive for \emph{usefulness} (Q1), \emph{coherence/connection to prior chats} (Q7), \emph{motivation to follow up} (Q9), and \emph{willingness to reuse} (Q10): a majority of participants selected \emph{Agree} or \emph{Strongly Agree} on these items. Second, several items show mixed responses: \emph{trust in correctness} (Q8), \emph{lesson length/depth matching users' time budgets} (Q6), and \emph{degree of repetition} (Q2; reverse-coded in the figure) received a spread of answers (neutral to agree/disagree), indicating variability in perceived content quality, granularity, and redundancy across participants. Items about \emph{novelty} and \emph{gap identification} (Q4, Q5) tended toward agreement, suggesting LOOM often surfaces useful “unknowns” or new material for users.
% TODO: The labels here could be better aligned with the figure, e.g., add the labels "novelty" to each question, or we can annotate the graph with Q1-Q11 and refer to them that way.

\subsubsection{Qualitative themes:} Analysis of the nine free-text responses revealed consistent themes that help explain the quantitative patterns.

\textbf{What participants liked most:} Participants commonly praised the modular lesson structure and short quizzes for helping consolidation and recall (e.g., P2: ``modularized flash cards with small quizzes really make me feel like I understand the knowledge''). Several respondents emphasized perceived personalization and coherence with prior chats (P7: ``the lessons feel personal and sometimes give me things that i wouldn't have thought of, and relevant''). Others appreciated the visible trace of progress from grouped/completed lessons (P5: ``I like the grouping of completed lessons, shows what I learned'').

\textbf{Suggested improvements:} Participants requested stronger support for continuation and grouping of learning trajectories (P1: ``Group the suggested lessons; allow continuation within a group/existing lessons''), better filtering or weighting to prioritize instruction-focused sessions, and more diverse content formats (graphics, multimedia) and optional quiz controls (P7: ``I don't always want to do a quiz -- kinda unnecessary sometimes''). A few noted variable lesson quality and occasional redundancy between lessons (P8: ``Quality of lessons isn't always great, sometimes not detailed''; P9: ``Some lessons are too similar so seem redundant''). One participant reported slower response times with certain models, which affected perceived fluidity.

% \subsection{Discussion and limitations}

% \textbf{Interpretation:} This formative evaluation indicates that LOOM’s chat-to-course pipeline produces suggestions perceived as relevant and coherent for many users, and that the modular lessons and quizzes are valuable for consolidation. At the same time, mixed responses on trust, lesson granularity, and repetition highlight concrete opportunities: (1) improve content consistency and factual reliability, (2) tune lesson length and granularity to better match user time budgets, (3) reduce redundancy via stronger de-duplication/grouping logic, and (4) offer user controls for quiz frequency and content filters.

% \textbf{Limitations:} The study is formative with a small convenience sample ($n=10$) and short deployment window (approximately two days per participant). Outcomes are primarily self-reported perceptions rather than objective learning gains. The sample and study duration limit generalizability and preclude claims about long-term learning effects.

% \textbf{Next steps:} We plan to incorporate participant suggestions (grouped continuations, improved filtering, optional quizzes, richer media) and to run a larger, longitudinal study with objective pre/post measures of knowledge on targeted topics. We will also formalize and report the graph-construction and overlap-deduplication methods used to generate course content, and evaluate alternative approaches in an ablation study.

\section{Discussion}

\subsubsection{Interpretation}
Our formative study suggests that \sys{} can turn everyday LLM chats into structured mini-courses that many participants described as relevant, coherent, and personally meaningful. Participants often recognized their own phrasing and problems in the generated lessons, and several highlighted that short quizzes and grouped progress “made it feel like I was actually learning rather than just asking one-off questions.” This supports our core claim: that an assistant can notice recurring intent in spontaneous conversations and respond with guidance toward mastery, not just answers in the moment.

At the same time, responses revealed a persistent tension between \textit{initiative} and \textit{continuity}. Participants generally liked that \sys{} proactively surfaced “things I wouldn’t have thought to ask,” but they also reported uneven depth, occasional redundancy, or lessons that did not match their available time or immediate goals. In other words, \sys{} sometimes behaved like a tutor advancing a longer-term trajectory, and sometimes like an assistant interrupting with content the learner was not ready to invest in. This validates the design space but also clarifies a requirement: proactive guidance must stay aligned with what the learner can actually act on \textit{right now}, or it risks becoming noise. Balancing opportunistic initiative (act in the moment) with credible continuity (build toward mastery) is therefore not just a systems problem, but an interaction contract problem.

\subsubsection{Limitations}
This study is formative. The sample was small ($n=10$), recruited via convenience sampling, and each participant only used the system for roughly two days without any comparative baselines. Our measures emphasize perceived usefulness, coherence, and motivation rather than objective learning gains or retention. We did not include pre/post testing, spaced review, or delayed recall, so we cannot yet claim that \sys{} improves durable understanding.

Technically, the current prototype relies heavily on prompting foundation models to (i) summarize chats, (ii) propose outlines, (iii) generate lesson content and quizzes, and (iv) update the learner memory graph. Participants noticed the consequences: variable factual confidence, inconsistent granularity across lessons, and repeated or overlapping topics. Since the learner memory graph is also updated through LLM inferences, the system can overgeneralize what the learner “knows” after a single interaction. These limitations cap both reliability and trust.

\subsubsection{Future Plan}
We outline two main directions: improving the robustness of the pipeline itself, and deepening learner-facing interaction with the system’s internal model.

\textbf{(1) Pipeline and agent robustness.}
Right now, most stages of \sys{} are driven by end-to-end prompting. Our results suggest that this is workable for generating plausible lessons, but too fragile for sustained use. We plan to move toward a more auditable, modular pipeline with explicit intermediate representations.

Instead of asking an LLM to directly “write a course,” \sys{} will first generate structured plans: learning objectives, prerequisite concepts, misconceptions to address, and assessment checkpoints. These plans can then be validated or templated before realization as final lesson text. Quizzes can be generated under lightweight constraints (e.g., one unambiguous correct option, no duplicated distractors) and optionally checked against small curated item banks. This separation of \emph{planning} from \emph{surface text} is intended to increase pedagogical consistency, reduce redundancy across lessons, and improve factual reliability.

We also plan to harden the learner memory graph using signals beyond LLM inference alone. Instead of treating model guesses as truth, the graph will incorporate observable evidence: frequency and recency of related queries, explicit self-reports (“I already know this,” “I’m confused here”), and quiz performance. This should let the system distinguish “I troubleshot this once at work” from “I am actively studying this,” and avoid prematurely marking concepts as mastered. Together, these steps aim to make \sys{} more trustworthy, inspectable, and less brittle than a pure prompt chain.

\textbf{(2) Interaction and learner control.}
Participants asked for clearer continuation and more fine-grained control over how lessons evolve. In the current prototype, \sys{} already visualizes the goal umbrellas and allows learners to mark modules as “already known,” giving them some influence over what counts as progress. However, these actions are still largely reactive: learners can respond to what the system proposes but cannot directly manipulate or extend the structure itself.

Future iterations will deepen this into a \textit{mixed-initiative interaction model}, where both learner and system can propose and negotiate updates to the learner graph. For example, learners could manually create new goals (“start a track on data ethics”), merge or reprioritize existing umbrellas, or request a new module that bridges two visible areas of their graph. Conversely, the system could surface candidate connections—such as suggesting that two goal areas overlap conceptually or that a recently discussed topic fits under an existing goal—and let the learner confirm or reject them. This two-way control turns the learner memory graph into a shared workspace rather than an opaque record.

We also plan to contextualize the visualization to show \textit{relations among goals}, not just their internal modules, helping learners see how areas of expertise connect and evolve over time. Together, these interaction improvements aim to balance system initiative with learner agency: the system can still act opportunistically, but always in a way that is legible, editable, and grounded in the learner’s own framing of their goals. Longer term, we will study how such mixed-initiative curation affects engagement, trust, and sustained learning outcomes in extended deployments.

\section{Conclusion}
LOOM explores how everyday interactions with LLMs can evolve into sustained, personalized learning trajectories. By combining conversational observation, adaptive content generation, and a dynamic learner memory graph, LOOM unifies continuity and initiative in a single agentic pipeline. Our formative study with ten participants suggests that such chat-to-course transformation can yield learning materials perceived as relevant, coherent, and personally meaningful, while also revealing needs for stronger content consistency and user control.
Looking forward, we aim to strengthen LOOM’s pipeline through more modular, auditable agents and to advance mixed-initiative interaction, enabling learners to directly shape and extend their own learning graphs. These next steps will help transform reactive chat assistants into long-term learning companions that are both proactive and pedagogically grounded.

\bibliography{aaai2026}


\end{document}